\documentclass[showpacs,aps,twocolumn]{revtex4}
\usepackage{graphicx}
\usepackage{amsfonts}
\usepackage{amsmath}
\usepackage{amssymb}
\usepackage{multirow}
\usepackage{braket}
\usepackage[usenames,dvipsnames]{color}

\begin{document}
\bibliographystyle{apsrev}

\title{Continuous stochastic processes with non-local memory}

\author{S.~S.~Melnyk}
\affiliation{A. Ya. Usikov Institute for Radiophysics and Electronics NASU, 61085 Kharkiv, Ukraine}

\author{V.~A.~Yampol'skii}
\affiliation{A. Ya. Usikov Institute for Radiophysics and Electronics NASU, 61085 Kharkov, Ukraine}
\affiliation{V. N. Karazin Kharkov National University, 61077 Kharkov, Ukraine}

\author{O.~V.~Usatenko}
\affiliation{A. Ya. Usikov Institute for Radiophysics and Electronics NASU, 61085 Kharkov, Ukraine}
\affiliation{V. N. Karazin Kharkov National University, 61077 Kharkov, Ukraine}

\begin{abstract}
We study the non-Markovian random continuous processes described by the Mori-Zwanzig equation. As a starting point, we use the Markovian Gaussian Ornstein-Uhlenbeck process and introduce an integral memory term depending on the past of the process into expression for the higher-order transition probability function and stochastic differential equation. We show that the proposed processes can be considered as continuous-time interpolations of discrete-time higher-order auto-regressive sequences. An equation connecting the memory function (the kernel of integral term) and the two-point correlation function is obtained. A condition for stationarity of the process is established. We suggest a method to generate stationary continuous stochastic processes with prescribed pair correlation function. As illustration, some examples of numerical simulation of the processes with non-local memory are presented.
\end{abstract}


\maketitle

\section{Introduction}

Nature offers us a large number of examples of random processes. They occur in the science just so often as deterministic ones. A systematic understanding of these processes is necessary to describe a vast range of complex phenomena. However, no such general construction currently exists. The Markov processes are the simplest and the most popular examples for describing the random phenomena (see, e.g.,~\cite{Kampen,gar,Horsthemke}). The two Gaussian Markov processes, Brownian motion and Ornstein-Uhlenbeck (OU) process~\cite{Uhlen}, have been used extensively in various applications from financial mathematics to natural sciences. It is important to note that these processes use the local memory of the past known event in their history. Indeed, in the Ornstein-Uhlenbeck model, the transition probability density $\rho$ for the random variable $V$ to take on the value $v$ at the moment $t+\tau$ of time depends on the last known point $V(t)=v_1$ only,
\begin{eqnarray}\label{-1}
&&\rho \big[V(t+\tau)=v|V(t)=v_1\big] \nonumber\\[6pt]
&&=\sqrt{\frac{\nu}{\pi \sigma^2 (1-\textrm{e}^{-2 \nu \tau})}}
\!\exp\left[-\frac{(v-v_1\textrm{e}^{-
\nu\tau})^2}{\sigma^2(1-\textrm{e}^{-2 \nu\tau})/\nu}\right]\!.
\end{eqnarray}
Here $\nu$ and $\sigma$ are two independent parameters of the model.

For any $\nu >0$, the conditional probability distribution function (CPDF) \eqref{-1} provides the stationarity of the OU process with the Gaussian one-point distribution function $\rho_1 (V(t)=v)=(1/{\sqrt{\pi \sigma^2/\nu}}) \exp(-{\nu v^2/\sigma^2})$. The limiting case $\nu =0$ corresponds to the non-stationary Brownian motion. The correlation function of the OU process is proportional to $\exp(-\nu \tau)$,
\begin{eqnarray}\label{0}
C(\tau)=\langle V(t+\tau) V(t)\rangle = \frac{\sigma^2}{2\nu}\exp{(-\nu\tau)},
\end{eqnarray}
where the symbol $\langle ... \rangle$ denotes a statistical average.

Equation \eqref{-1} defines the Markovian process because it contains the local ``memory'' of its past at only one temporal point $t$.

The Ornstein-Uhlenbeck process can be described by the Langevin equation or, more correctly from the mathematical point, by the
stochastic differential equation (SDE)~\cite{gar},
\begin{eqnarray} \label{-2}
dV(t)=-\nu V(t) dt + \sigma \, dW(t).
\end{eqnarray}
Here $dW(t)$ is the standard white noise, i.e., $W(t)$ is the continuous centered Wiener process with independent increments with variances $\langle (W(t+\tau)-W(t))^2\rangle = |\tau|$, or, equivalently, $W(t)= \int dW(t) \Rightarrow $ $\langle
dW(t)dW(t')\rangle =\delta(t-t')dtdt'$. The term $-\nu V(t) dt$ in Eq.~\eqref{-2} describes a local one-point feature of the process. The positive value of the constant $\nu$ provides an anti-persistent character of the process with attraction of $V(t)$ to the point $V=0$. As seen from Eq.~\eqref{-2}, the Ornstein-Uhlenbeck process simulates the Brownian motion of a microscopic particle in a liquid viscous suspension subjected to a random force with intensity $\sigma$.

According to the Doob theorem (see, e.g., Ref.~\cite{Kampen} and references therein), the OU process is the unique continuous
Markovian stationary Gaussian process.

The classical books on probability and stochastics hardly mention the non-Markov processes despite they are most general in nature. A lot of systems in the real world are more complex than the Markovian ones, they have non-Markovian character of the memory (see, e.g.,~\cite{mok,bre,sie,ros,sta}). Therefore, it is necessary to go beyond the simple Markovian model based on Brownian motion and Ornstein-Uhlenbeck process. There is a huge literature on various kinds of non-Markovian processes. To begin with, the so-called semi-Markovian processes provide a very broad and popular class (see, e.g., Refs.~\cite{Schlesinger, Scher, Hughes, Bouchaud, Metzler2, Goychuk4, Goychuk6}). Note that, in recent years, a lot of attention has been paid to the study of non-Markov processes due to their role in decoherence phenomena in open quantum systems (see, e.g., Refs~\cite{lam, bre, kan}). Namely, non-Markovianity can serve as a source for suppressing the exponential decay of coherence in the interaction of a quantum system with a classical thermal bath~\cite{bel,chi,byl}.

As has been mentioned by many authors (see, e.g., Refs.~\cite{Hanggi1, Cox, Kampen}), all non-Markovian processes are
``history''-dependent. In this paper, we take into account this history in the explicit form introducing an integral non-local
memory term into Eq.~\eqref{-2},
\begin{eqnarray} \label{12}
dV(t)&=&-\nu V(t) dt \\[6pt]
 & - &\int_{0} ^{\infty} \mu(t')
V(t-t')dt'dt +\sigma \, dW(t),\nonumber
\end{eqnarray}
where $\mu(t)$ is the so-called memory function.

Such a generalization of SDE has been discussed by many authors~\cite{Adelman, Hanggi_Thomas, Hynes, Wang, Goychuk1}. In most cases, the so-called internal noise was considered, when, according to the fluctuation-dissipation theorem~\cite{Kubo}, the function $\mu(t)$  is uniquely determined by the correlator of the stochastic perturbation $W(t)$. Then the memory kernel $\mu(t)$ describes the so-called viscoelastic friction~\cite{Goychuk1}. However, in the case of external noise, the fluctuation and dissipation come from different sources, and the frictional kernel $\mu(t)$ and the correlation function of the noise are independent (see, e.g., Ref.~\cite{Wang}). In the present paper, we consider an \emph{arbitrary} memory kernel $\mu(t)$ and the Gaussian \emph{external} noise $W(t)$ \emph{independent} of $\mu(t)$. Equation~\eqref{12} could be a good physical model for the systems where the external noise is much more intensive than the thermal one. 

After introducing the integral memory term, Eq.~\eqref{-2} becomes the stochastic integro-differential Mori-Zwanzig equation~\eqref{12}, see Refs.~\cite{Mori, Zwanzig1, Nordholm, Kawasaki, Grabert}. Note that it is also a particular case of the generalized Langevin equation (GLE) without deterministic external force. The GLE describes various very important physical processes, from fractional Brownian motion to stochastic hydrodynamic memory effects resulting in algebraic tails of velocity autocorrelation functions (see, e.g., Refs.~\cite{Bogolyubov, Ford, Kubo1, Zwanzig1, Kubo, Zwanzig, Weiss}).

For the \emph{infinitesimal} time intervals $dt$, SIDE \eqref{12} engenders the \emph{higher-order} transition probability function of the form,

\begin{widetext}
\begin{eqnarray}\label{11}
  \rho\big[V(t+dt) = v|V(t'\leqslant t)=v(t')\big] = \frac{1}{\sqrt{2\pi \sigma^2 dt}} \exp\left[-\frac{[v-v(t)(1 - \nu dt)+ dt \int_0^\infty \mu(t') v(t-t')dt']^2}{2\sigma^2 dt}\right].
\end{eqnarray}
\end{widetext}
This formulae cannot be considered as a generalization of the Ornstein-Uhlenbeck one because Eq.~\eqref{-1} is valid for any $\tau$, not only infinitesimal. For the OU process, CPDF~\eqref{-1} describes the transition from the last known point of the past to the current time whereas the transition probability~\eqref{11} takes into account the whole past history of the process. Just by this reason, stochastic Eq.~\eqref{12} becomes integro-differential, and the process with memory $\mu (t)$ is non-Markovian by definition.

It is important to note that the left-hand side of expression~\eqref{11} is not the conditional probability distribution density of $V(t+dt)$ for known $V(t')$ at fixed point $t'$. Actually, it describes the dependence of the distribution $V(t+dt)$ on the whole history of the process $V(t')$ at times $-\infty<t'<t$, which ensures non-Markov properties of the process. In other words, the higher-order distribution function~\eqref{11} is a functional of the history $V(t')$ with $-\infty <t'<t$.

In principle, using Eq.~\eqref{12}, one can derive the two-time conditional probability density  for arbitrary $\tau$. However, this conditional probability, contrary to Eq.~\eqref{11}, is not suitable for definition of the process under consideration.

Emphasize that both Eqs.~\eqref{12} and \eqref{11} equally allow us to construct a sample of the process. Indeed, the right-hand side of Eq.~\eqref{11} represents the Gaussian function with the characteristic width $\sigma \sqrt{dt}$ and centered at the point $v(t)-\nu v(t) dt  - \int_{0} ^{\infty} \mu(t') v(t-t')dt' dt$. At infinitesimal $dt$, this imposes for the differential $dv = v - v(t)$, occurring during time $dt$, to consist of two terms, deterministic and stochastic ones. The deterministic history-dependent term is $-\nu v(t) dt - \int_{0} ^{\infty} \mu(t') v(t-t')dt'dt$ that coincides with the first two terms in the right-hand side of stochastic integro-differential equation (SIDE)~\eqref{12}. Thus, Eq.~\eqref{11} can be rewritten in the form,
\begin{widetext}
\begin{eqnarray}\label{11a}
  \rho\big[V(t+dt) = v|V(t'\leqslant t)=v(t')\big] = \frac{1}{\sqrt{2\pi \sigma^2 dt}} \exp\left[-\frac{[dv/dt + \nu v + \int_0^\infty \mu(t') v(t-t')dt']^2 dt}{2\sigma^2}\right],
\end{eqnarray}
\end{widetext}
where the numerator of the exponent contains explicitly the deterministic part of SIDE \eqref{12}.

Further our analytical study is based on Eq.~\eqref{12}. We give Eqs.~\eqref{11} and~\eqref{11a} only for the clarification of the relation between the model under consideration and the Ornstein-Uhlenbeck model. In addition, it clarifies the method of numerical modeling for such systems. When constructing a concrete realization of the process, we can use equally Eq.~\eqref{12} or Eq.~\eqref{11}. In the first case, we calculate the deterministic term on the base of the already occurred history of the process and then add the random term. When using Eq.~\eqref{11}, we  straight away generate a random term in the vicinity of deterministic one. An example of program code that solves this problem of numerical construction of random processes is given by Ref.~\cite{Program}.

There exists a direct analogy between the processes considered here and well-known discrete high-order auto-regressive sequences (see, e.g., Ref.~\cite{Arratia,Yule,Walker}). Such sequences are also characterized by non-local memory. We pay a special attention to the existence of important relation between the memory function and two-point correlation function, which is valid for auto-regressive sequences~\cite{Yule,Walker}. Below we derive the similar relation for the continuous-time process, Eqs.~\eqref{12} and~\eqref{11}, with non-local memory. This relation allows one to construct a continuous Gaussian process with any prescribed two-point correlation function. We demonstrate this ability by numerical simulations using two simple model examples of the memory functions presenting different kinds of correlations. In conclusion we discuss some possible generalizations of the considered construction.

\section{Correlation function}

In this section, we show that the process \eqref{12}, \eqref{11} can be considered as a continuous-time interpolation of the known discrete-time higher-order auto-regressive (AR) sequence with similar statistical properties. The transition to continuous regime will allow us to generalize the Yule-Walker relation between the correlation and memory functions valid for AR sequences to the case of continuous processes with non-local memory.

The auto-regressive $N$-th order model, describing a random discrete-time sequence $V_n$ with the memory extended to $N$
previously realized terms, is
\begin{equation}\label{AM_def}
V_n = \sum_{r=1}^N f_r V_{n-r} + \sigma \Delta W_n (\Delta t),
\end{equation}
where $f_r$ is the discrete memory function, $\Delta W_n (\Delta t) = W_n (\Delta t) - W_{n-1} (\Delta t)$ is the discrete white noise with $n$-independent variances $ \langle (W_n(\Delta t)-W_{n-1}(\Delta t))^2\rangle = \Delta t$. It is suitable to put
the sequence \eqref{AM_def} onto the temporal axis with the time interval $\Delta t$ between $n$-th and $(n-1)$-th events, i.e.,
$V_n = V(t_n) = V(n \Delta t)$. Then we present the local and non-local memory terms, $f_1$ and $f_r\, (r\geqslant 2)$, in the form
\begin{equation}\label{f}
f_1 = 1 - \nu \Delta t,\,\, r=1;\,\, \quad f_r = -\Delta t^2\mu(r \Delta t), \,\,
r=2,3,\ldots
\end{equation}
In the limit $\Delta t \rightarrow dt$, $\Delta t^2 \rightarrow dtdt'$, Eq.~\eqref{AM_def} transforms into Eq.~\eqref{12}.

The memory coefficients $f_r$ for the auto-regressive sequences are closely related to the correlation function $C_n$ by the Yule-Walker equation~\cite{Yule,Walker},
\begin{equation}\label{C_n}
C_n = \sum_{r=1}^\infty f_r C_{n-r}, \quad n>0.
\end{equation}
The continuous Yule-Walker relation, following from Eqs.~\eqref{f} and~\eqref{C_n}, is
\begin{equation}\label{AM_C_Phi} \frac{d C(t)}{dt} + \nu C(t) +
\int_0^\infty \mu(t') C(t - t') dt' =0, \quad t > 0.
\end{equation}
This equation can be obtained not only by the limiting transition $\Delta t \rightarrow 0$, but also by means of SIDE~\eqref{12}. Multiplying Eq.~\eqref{12} by $V(t-\tilde t\,), (\tilde t\,>0)$ and averaging it over the statistical ensemble, we have
\begin{equation}
C(\tilde t\,) - C(\tilde t\,-dt)+ \nu C(\tilde t\,) dt + dt
\int_0^\infty dt' \mu(t') C(\tilde t\, - t')  = 0.
\end{equation}
Dividing this equation by $dt$, we get Eq.~\eqref{AM_C_Phi}.

The continuous Yule-Walker equation defines the correlation function up to unknown normalization factor $C(0)$. The normalization condition can be derived with the use of the time-independence of the one-point variance $\langle V^2(t) \rangle = C(0)$. Taking into account this condition in the form $\langle V^2(t+dt) \rangle = \langle V^2(t)\rangle$ with $V(t+dt) \approx V(t) + dV(t)$, neglecting the terms proportional to $(dt)^2$ in $\langle V^2(t+dt) \rangle$ and using Eqs.~\eqref{12} and \eqref{AM_C_Phi}, after some simple algebraic operations, we get the following relation:
\begin{equation}\label{nc}
    \frac{dC(t)}{dt}{\Big|_{t=0_+}} = -\frac{\sigma^2}{2}.
\end{equation}
Here the subscript ``$+$'' shows that the derivative is taken at positive $t$ close to zero. Note that this equation is valid to describe not only processes with non-local memory, but also with the local one, see Eq.~\eqref{0}.

Thus, Eqs.~\eqref{AM_C_Phi} and \eqref{nc} being supplemented by equation
\begin{equation}\label{Cor_is_pair}
     C(-t)=C(t)
\end{equation}
form the complete set of equations. They allow one finding the correlation function $C(t)$ if the parameters $\sigma, \nu$ and $\mu(t)$ are known, or to find $\nu$, $\mu(t)$ and $\sigma$ for known $C(t)$.

Equations \eqref{AM_C_Phi}, \eqref{nc} and \eqref{Cor_is_pair} represent a very useful tool for studying the statistical properties of random processes with non-local memory. These properties are governed by the constants $\nu$ and $\sigma$ and the memory function $\mu(t)$. To avoid some complications, we will suppose that this function, $\mu(t)$, has good properties at $t\rightarrow\infty$. More exactly, we assume that the function $\mu(t)$ has either a finite characteristic scale $T$ of decrease, or it abruptly decreases to zero at $t>T$, $\mu(t>T)=0$.

Equations~\eqref{AM_C_Phi}, \eqref{nc} and \eqref{Cor_is_pair} enable one to solve two kinds of problems. First, for given $\nu$, $\sigma$, and $\mu (t)$, solving integro-differential equation ~\eqref{AM_C_Phi} with respect to $C(t)$, one can find the correlation function of the process, even without numerical construction of the process. The inverse problem of finding $\nu$, $\sigma$, and $\mu (t)$ for given $C(t)$ is much more important and interesting. We are able to construct a continuous Gaussian process with any prescribed two-point correlation function, satisfying natural restrictions $C(t)< C(0), C(t\rightarrow\infty)\rightarrow 0$. Using Eqs.~\eqref{AM_C_Phi} and~\eqref{nc}, we can find numerically the parameters $\nu$, $\sigma$ and the memory function $\mu (t)$ that provide the desired correlator $C(t)$. This means that we can determine these characteristics for any natural continuous non-Markovian process.

\section{Stationarity}

In this section, we analyze the conditions imposed on $\nu$ and function $\mu(t)$ necessary for stationarity of the considered here process with non-local memory.

The condition of the stationarity can be examined by an analysis of asymptotical properties of the correlation function. We carry out this analysis for processes with finite depth $T$ of the memory function, when $\mu(t>T)$ = 0. In this case, the correlation function can be presented as a sum of exponential terms of the form,
\begin{eqnarray} \label{101}
A\exp (-z t / T),
\end{eqnarray}
for $t \gg T$. Equation~\eqref{AM_C_Phi} yields  the following characteristic algebraic equation for the complex damping coefficients $z$:
\begin{eqnarray} \label{102}
z/ T=\nu +  \int_{0}^T \mu(t)\exp (z t/T) d t.
\end{eqnarray}
For the  process to be stationary, all roots $z_i$ of this equation should obviously have real positive parts, $\Re \,z_i >0$. Consider
Eq.~\eqref{102} for $\nu =0$. In this case, the stationarity of the process can be provided by the positive memory function, $\mu(t) >0$. However, contrary to the local memory in the OU process, where any $\nu>0$ provides the stationarity, the non-local positive memory cannot always do this. To demonstrate this, we consider two simple models of the memory functions.

First example is the local, but remote from the instant time moment $t$, memory function,
\begin{eqnarray} \label{103}
\mu (t) = \frac{\mu _0}{T} \delta (t-T),
\end{eqnarray}
where $\delta$ denotes Dirac's delta.  In this case, and at $\nu=0$, Eq.~\eqref{102} gives
\begin{eqnarray} \label{104}
z=\mu_0 \textrm{e}^z.
\end{eqnarray}

We are interested in the root of this equation with the lowest real part $(\Re \, z)_{\textrm{min}}$ because the exponential function in Eq.~\eqref{101} with this exponent defines behavior of the correlation function at $t \rightarrow \infty$.

In the interval $0<\mu_0 <\mu_{\textrm{c}}=1/\textrm{e}$, there exists an infinite set of the pairs of complex-conjugated roots of Eq.~\eqref{104}. In addition, there exist  two real positive roots, one of which is less than the real parts of all other roots. This means that the asymptotic behavior of the correlation function in Eq.~\eqref{101} is exponential. Within the
interval $1/\textrm{e}<\mu_0 < \mu_{\textrm{crit}} = \pi/2$, the real roots do not exist, i.e., all roots have nonzero imaginary parts. Therefore, the exponential decrease of the correlation function is accompanied by oscillations. At last, for high enough values of $\mu_0$, at $\mu_0 > \mu_{\textrm{crit}} $, the roots of Eq.~\eqref{104} with negative real parts appear. This means that the process becomes non-stationary even for positive memory functions.

The second example is the step-wise memory function~\cite{UYa,RewUAMM},
\begin{eqnarray} \label{105}
\mu (t) = \frac{\mu _0}{T^2} \theta (T-t),
\end{eqnarray}
where $\theta(x)$ is the Heaviside theta-function. In this case, Eq.~\eqref{102} gives
\begin{eqnarray} \label{106}
z^2=\mu_0 (\textrm{e}^z-1).
\end{eqnarray}
For $0<\mu_0 <\mu_{\textrm{c}} \approx 0.648$, the root of Eq.~\eqref{106} with minimal real part does not have imaginary part, that corresponds to exponential decay of the correlation function. In the interval $\mu_{\textrm{c}}<\mu_0 < \mu_{\textrm{crit}} = \pi^2/2$, all roots of Eq.~\eqref{106} are complex with positive real parts and nonzero imaginary parts, i.e., the damping of the correlation function is accompanied by oscillations. At $\mu_0 > \mu_{\textrm{crit}}$, the roots of Eq.~\eqref{106} with negative real parts appear. This means that the process with the step-wise memory function becomes also non-stationary at high values of $\mu_0$.

Thus, we have come to a quite unexpected result. Contrary to the OU process with local memory, where the only one limiting point $\nu = 0$ separates the stationary and non-stationary regimes, there exist two separating values of $\mu_0$ for the process with non-local memory. This means that the stationary regime can be realized for $\mu_0$ within some finite interval $0<\mu_0 < \mu_{\textrm{crit}}$ of the memory amplitude. At  bottom limiting point, $\mu_0 = 0$, we transit to the case of trivial Brownian motion while at upper limiting point, $\mu_0 \rightarrow \mu_{\textrm{crit}}$, we have the correlated Brownian motion with an oscillating correlation function.

\begin{figure}[h!]
\center\includegraphics[width=0.5\textwidth]{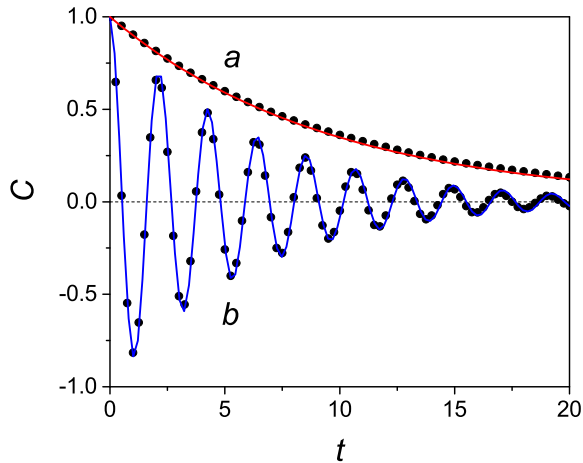}
\caption{(Color online) The correlation function of the process with step-wise memory function Eq.~\eqref{105} at different values of its amplitude, $\mu_0 = 0.1$ (a) and $\mu_0 = 4.0$ (b). Black filled circles present the results of numerical generation of the process using the higher-order conditional probability~\eqref{11} whereas the solid lines are plotted using the analytical formulas. Parameters of the generated processes: $\nu = 0$, $T = 1.0$, $\Delta t = 0.025$. The value of $\sigma$ is chosen to provide the equality $C(0) = 1$.} \label{figGen1}
\end{figure}

\section{Numerical simulations}

In this section, we demonstrate our ability to numerically generate the random process with given parameters $\nu$, $\sigma$ and memory function $\mu(t)$, and to solve the direct problem of finding the correlation function. We can also solve the inverse problem of generating the random process with prescribed correlation function $C(t)$. Solving the inverse problem implies finding the parameters $\nu$, $\sigma$ and memory function $\mu(t)$ for generation of this process.

When generating the process numerically, one can arbitrarily specify the values $V(t)$ of the initial part of sequence, at $0<t<T$, since a stationary process after a sufficiently long time $t\gg T$ becomes independent of them.

Note that the stationarity of a Gaussian process together with a decay of correlations, i.e. \emph{$C(t \to \infty)=0$}, leads, according to Ref.~\cite{pap}, to the ergodicity in correlations. This property is very useful for numerical calculations since the averaging procedure can be done over the time of the process, and the ensemble averaging can be avoided. Therefore, we always apply averaging over $t$ in our numerical calculations.

In the numerical solution of the direct and inverse problems, we always deal with discrete time. The time step $\Delta t$ is chosen to be much smaller than all other characteristic time scales of the model, i.e., $\nu \Delta t \ll 1$, $\mu T\Delta t  \ll 1$. After selection of the parameters $\nu$, $\sigma$ and function $\mu(t)$, generation of the process by Eqs.~\eqref{AM_def} and~\eqref{f} can be performed (the direct problem).

As mentioned above, both Eqs.~\eqref{12} and \eqref{11} equally allow us to construct a sample of the process. In fact, we use expression~\eqref{12} for the generation of random process.

Similarly, knowing the prescribed correlation function $C(t)$, using the discrete representation $C_n = C_{t/\Delta t} =C(t)$, we can solve the Yule-Walker set of linear equations~\eqref{C_n} by the convenient numerical methods (Cramer's, Gaussian elimination, iterative, or others) for the unknown parameters $\nu, \,f_r, \, \sigma$ (the inverse problem) and then to construct a process.

The course of numerical solution of both problems, direct and inverse, is described in more detail in the Appendix.

\begin{figure}[h!]
\center\includegraphics[width=0.5\textwidth]{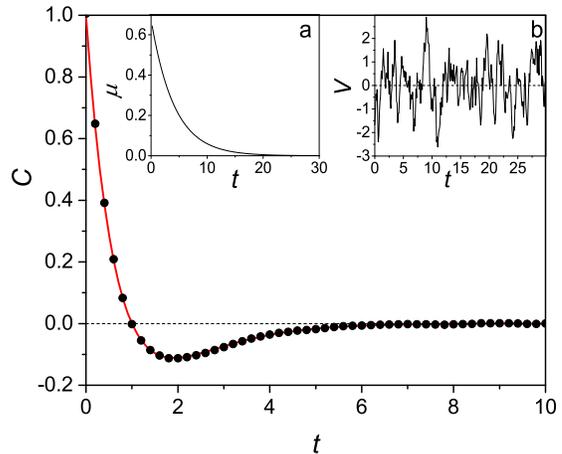}
\caption{(Color online) The main panel: the prescribed correlation function, $C(t) = (1-t) \exp (-1.1 t)$, (red solid line) and numerically calculated one (black filled circles). Inset (a): the estimated memory function $\mu(t)$. Inset (b): the fragment of the generated random process with the memory function shown in the inset (a). The parameters: overall time of the process is $10^5$, time step $\Delta t = 0.1$, and the estimated parameter $\nu = 1.9$.} \label{figGen2}
\end{figure}

In Fig.~\ref{figGen1}, we present the results of numerical generation of the process with given parameters $\nu$, $\sigma$ and
memory function $\mu(t)$, as well as the correlation function for the constructed process. We choose the step-wise memory function, Eq.~\eqref{105}, with $\mu_0 = 0.1$ and $\mu_0 = 4.0$ and constructed numerically the correlated sequence using the higher-order conditional probability~\eqref{11}, see Ref.~\cite{Program}. The first value of $\mu_0 $ is chosen within the interval $0<\mu_0 <\mu_{\textrm{c}} \approx 0.648$ where the correlation function has the exponential asymptotics with positive value of $z$, 
\begin{eqnarray} \label{as1}
C \simeq \exp (-0.105 \,\,t/T),
\end{eqnarray}
see the red solid curve. The second value of $\mu_0$ is within the interval $\mu_\textrm{c} <\mu_0 <\mu_{\textrm{crit}} = \pi^2/2$ where the damping of $C(t)$ is accompanied by the oscillations, 
\begin{eqnarray} \label{as2}
C \simeq \cos(2.94 \,\,t/T) \exp (-0.162 \,\,t/T) ,
\end{eqnarray}
see the blue solid curve. The damping constants and frequency in Eqs.~\eqref{as1} and \eqref{as2} are the numerically obtained solutions of Eq.~\eqref{106} with minimal real parts for the parameters $\mu_0$ indicated above. It is noteworthy that analytical asymptotics Eqs.~\eqref{as1} and \eqref{as2}, which contain only one exponential term, approximate well the results of numerical analysis even for small values of $t$.

In Fig.~\ref{figGen2}, we demonstrate our ability to solve the inverse problem. The red solid line in the main panel shows the prescribed correlation function $C(t) = (1-t) \exp (-1.1 t)$. The inset (a) presents the calculated memory function $\mu(t)$ which is close to the exponential function $0.64 \exp (-0.34 t)$, and the inset (b) is the fragment of the generated process with this $\mu(t)$ and $\nu = 1.9$. At last, the black filled circles in the main panel show the numerically calculated correlation function for the generated process.

\section{Conclusion}

Using the Markovian OU random process as the starting point and introducing an integral memory term depending on the past of the process into the expression for the higher-order transition probability function and stochastic differential equation, we study a class of continuous random processes with non-local memory. We show that these processes can be considered as continuous-time interpolations of discrete-time higher-order auto-regressive sequences. An equation, analogous to the Yule-Walker one, connecting the memory function (the kernel of integral term) and the two-point correlation function is obtained. This equation enables us to solve two kinds of problems -- the direct and inverse ones. We demonstrate an ability to numerically generate  the random process with given parameters $\nu$, $\sigma$ and memory function $\mu(t)$. We also examine the inverse problem of generating the random process with prescribed correlation function $C(t)$. Solving this problem implies finding the parameters $\nu$, $\sigma$ and memory function $\mu(t)$ for generation of this process. A condition for stationarity of the process is established. This condition is examined by an analysis of asymptotical properties of the correlation function. We give two examples of numerical simulation of the processes with non-local memory, demonstrating our ability to solve the direct and inverse problems.

We see several ways for generalization of the considered here model of processes with non-local memory: (i) to consider the processes with non-Gaussian (e.g., Levy-like) noise and/or with multiplicative noise; (ii) to study the processes with nonlinear and/or non-additive memory functions; (iii) to investigate the non-stationary Brownian motion with non-local memory; (iv) to consider the discontinuous processes (of the telegraph-like kind) with non-local memory. The particular attention should be paid to the application of the considered model for studying the non-local in time interaction of quantum systems with a heat bath (see, e.g., Ref.~\cite{Breuer}).

\appendix*

\section{}

We start the description of numerical procedure from the solution of the inverse problem, i.e., the construction of a discrete sample $V_n$ of a stochastic process $V(t)$ with a given correlation function $C(t)$ for $0<t<T$. The accuracy of the solution of the problem is determined by the discrete time step $\Delta t$. The sequence of operation is as follows:
\begin{enumerate}
\item Calculate the correlation function $C_n = C(\Delta t n)$ in the discrete set of points $n=0, \ldots N, \,\, N = T/\Delta t$;
\item Rewrite Eq.~\eqref{C_n} for the difference $\Delta C_n = C_{n+1}-C_n$ with summation limited by $N$ terms,
\begin{equation}\label{C_n_N}
\Delta C_n = - \sum_{r=0}^{N-1} \Delta t^2 \mu_r C_{n-r}, \quad n=0, \ldots N-1;
\end{equation}
here, instead of the parameter $\nu$, we introduce the zero-term $\mu_0 = \nu / \Delta t$;
\item Solve the set of $N$ linear equations~\eqref{C_n_N} with respect to unknown $\mu_r$ by one of the standard numerical methods;
\item In accordance with relation ~\eqref{nc}, find the value $\sigma = \sqrt{2(C_0 - C_1)/\Delta t}$;
\item Set the first $N$ values of $V$ to zero, $V_1 = V_2 = \ldots = V_N = 0$;
\item Using the calculated parameters of the process, construct iteratively the following values of $V$. For every $n \geqslant N$, the value of $V_{n+1}$ is generated with Gaussian probability density distribution with variance $\sigma^2 \Delta t$ and centered in the point
\begin{equation}\label{Xn1_center}
V_n - \sum_{r=0}^{N-1} \Delta t^2 \mu_r V_{n-r}, \quad n=N, \ldots;
\end{equation}
\item Continue the procedure until the required number of values $V_n$ has been constructed.
\end{enumerate}

The direct problem begins with the discretization of a given memory function, $\mu_n = \mu(n \Delta t), \,\, \mu_0 = \nu / \Delta t$, then we go to step 5 and perform the same follow-up actions. If the problem involves obtaining the specified standard deviation $\langle V_n^2 \rangle = C(0)$ , then, after construction, all values of $V_n$ should be proportionally normalized.

\end{document}